\documentclass{article}

\usepackage[accepted]{icml2026} 

\usepackage{microtype}
\usepackage{graphicx}
\usepackage{subcaption}
\usepackage{booktabs}
\usepackage{amsmath}
\usepackage{amssymb}
\usepackage{mathtools}
\usepackage{amsthm}
\usepackage{hyperref}
\usepackage{multirow}
\usepackage{tikz}
\usetikzlibrary{fit,backgrounds}
\usepackage{xurl}
\usepackage[table]{xcolor}

\usepackage[capitalize,noabbrev]{cleveref}

\theoremstyle{plain}

\theoremstyle{definition}

\theoremstyle{remark}

\newcommand{\refig}[1]{Fig.~\ref{#1}}
\newcommand{\refapp}[1]{App.~\ref{#1}}

\newcommand{\refsec}[1]{Sec.~\ref{#1}}

\usepackage{xcolor}

\usepackage{tcolorbox}
\newtcolorbox{hypothesisbox}[1]{colback=blue!5!white,colframe=blue!75!black,fonttitle=\bfseries,title=#1}

\usepackage{mathtools}

\usepackage{enumitem}
\definecolor{neublue}{HTML}{3F51B5}     
\definecolor{neugreen}{HTML}{2E7D32}    
\definecolor{neumagenta}{HTML}{AD1457}  
\definecolor{neupurple}{HTML}{8E24AA}   
\definecolor{neuorange}{HTML}{E64A19}   
\definecolor{neugreenpastel}{HTML}{A5D6A7} 
\definecolor{neulavender}{HTML}{E9D5FF} 

\icmltitlerunning{Against the Monolithic Wireless World Model: Why NextG Needs Composable and Agentic Intelligence}

\begin{document}

\twocolumn[
   \icmltitle{Against the Monolithic Wireless World Model: \\Why NextG Needs Composable and Agentic Intelligence} 

  \icmlsetsymbol{equal}{*}

  \begin{icmlauthorlist}
    \icmlauthor{Aladin~Djuhera}{TUM}
    \icmlauthor{Farhan~Ahmed}{IBM}
    \icmlauthor{Vlad~Andrei}{TUM}
    \icmlauthor{Swanand~Ravindra~Kadhe}{IBM}
    \icmlauthor{Alecio~Binotto}{Zeiss}
    \icmlauthor{Haris~Gacanin}{RWTH}
    \icmlauthor{Holger~Boche}{TUM}
  \end{icmlauthorlist}

  \icmlaffiliation{TUM}{Technical University Munich}
  \icmlaffiliation{RWTH}{RWTH Aachen University}
  \icmlaffiliation{IBM}{IBM Research}
  \icmlaffiliation{Zeiss}{Zeiss AG}

  \icmlcorrespondingauthor{Aladin~Djuhera}{aladin.djuhera@tum.de}

  \icmlkeywords{Machine Learning, ICML}

  \vskip 0.3in
]



\printAffiliationsAndNotice{}  


\begin{abstract}
    AI-native 6G visions increasingly invoke wireless foundation models, large multimodal models, and wireless world models as the natural endpoint of AI-native networking, drawing an analogy to recent developments in large language models (LLMs).
    We argue that this analogy is structurally incomplete.
    The success of LLMs is based on a broad, reusable, and largely self-contained tokenized data substrate, whereas the wireless domain lacks an equivalent data foundation.
    Unlike text, code, or images, wireless data such as CSI tensors, IQ samples, or scheduler logs are not self-contained: their meaning is configuration-dependent, simulator-conditioned, task-disaggregated, and weakly grounded in operational feedback, all structural bottlenecks that undermine current pre- and post-training recipes.
    We therefore argue that monolithic models, including mixture-of-experts (MoE) and wireless world models, are not the most realistic near-term path toward deployable AI-native networks.
    Instead, emerging evidence points toward composable and agentic network architectures, where general reasoning models orchestrate specialized signal processing models, classical algorithms, digital twins, standards-aware retrieval, and safety checks through explicit programmable interfaces.
\end{abstract}
\section{Introduction: Hype and Reality}
\label{sec:introduction}

AI-native visions of 6G and beyond describe a future in which learning systems permeate every layer of the wireless protocol stack, from physical-layer signal processing to network management and application-level semantics~\citep{saad2025agi, jiang2025comprehensive}.
This vision has catalyzed serious research into foundation models for several wireless applications, including channel estimation, task prediction, and sensing~\citep{djuhera2026mambacsphybridattentionstatespace,liu2025wifo,sheng2025wirelessfoundationmodelmultitask,yang2025generativeaimeetswireless}. 
Continuing this wave, many works are proposing so-called large wireless models (LWMs)~\citep{alikhani2025largewirelessmodellwm}, large multimodal models (LMMs)~\citep{LMMs}, and, more recently, wireless world models (WWMs)~\citep{zou2026telecom}, motivated by novel architectures like JEPA~\citep{assran2023self}. 
Such proposals advocate for training models that embed \emph{universal wireless intelligence} across all layers, similar as to how state-of-the-art LLMs seem to exhibit beyond-human-level intelligence across several diverse tasks.

\begin{quote}
\itshape But many years later, where is the ChatGPT equivalent to wireless foundation model intelligence?
\end{quote}

We argue that the LLM analogy is structurally incomplete and incompatible with the wireless protocol stack.
In particular, LLMs and world models alike work not only because of sheer model scale but, more importantly, because of broad, reusable, and self-contained tokenized data~\citep{kaplan2020scaling}.
However, wireless AI has neither ingredient in adequate form:
data samples are almost always fragmented across tasks, configurations, simulators, protocol layers, and operational contexts, distinguishing them sharply from text and code.
This is \emph{not} because the community has been insufficiently ambitious, but because of the high degree of modularity when designing wireless networks at scale.

In this paper, we make three main claims:
\begin{enumerate}[leftmargin=*]

    \item \textbf{The current AI-native direction is under-specified}: many works quickly adopt new AI architectures without defining compatible data interfaces, operational boundaries, or paths through existing telecom infrastructure. 
    
    \item \textbf{Wireless AI data suffers from four main bottlenecks}: configuration dependence, simulation and setup dependence, no universal wireless token, and missing operational feedback are detrimental to model development.

    \item \textbf{AI-native wireless must be agentic and compositional}: the protocol stack should not be replaced by monolithic all-purpose models, but instead, should combine agentic reasoning and tool calling with specialized models.
    
\end{enumerate}

We are not claiming wireless foundation models are useless or that LLMs have no role in wireless systems.
Our claim is narrower and constructive: the near-term path to AI-native NextG should be \emph{compositional rather than monolithic}, with \emph{agent harnesses} that orchestrate configuration-aware, provenance-annotated wireless data, specialized tools, and models through explicit interfaces.
\section{AI-Native NextG is Underspecified}
\label{sec:underspecified_AI}

The first wave of wireless AI produced valuable task-specific models. 
Neural networks have been successfully trained for channel estimation~\citep{jin2019channel}, beam prediction~\citep{li2023machine}, and resource allocation~\citep{djigal2022machine}, showing that wireless intelligence can be embedded in specific tasks and control loops. 
This is also reflected in the recent 3GPP Release-18/5G-Advanced work on AI/ML for the NR air interface~\citep{lin2023overview}, which emphasizes concrete use cases with implementation details and, most importantly, clear interface specifications for AI modules.

The current wave of AI aims to create foundation models that provide intelligence beyond the per-task level. 
This implies models that generalize across configurations, protocol settings, and simulation environments.
However, the operational meaning of such models is often vague or left underspecified.
\emph{What is the wireless analogue of a tokenized corpus?
Which observations, configurations, actions, and feedback signals should be included?
At which layer of the protocol stack should the model operate?}
Without answering these questions, ``AI-native'' risks becoming a label for applying generic post-training to telecom data, rather than a concrete architecture for deployable network intelligence.

Early telecom LLM efforts illustrate this limitation.
Fine-tuning models on standards documents and technical reports can improve domain question answering (Q\&A) and standards retrieval~\citep{TeleLLMs,TSpecLLM}, but they remain primarily language-level systems and thus cannot by themselves learn the dynamics of wireless channels or solve complex mobility or scheduling problems.
Furthermore, it still remains unclear how to measure actionable wireless intelligence due to the lack of adequate benchmarks.
Only recently have the GSMA Open-Telco LLM Benchmarks~\citep{gsma_open_telco_2025} included operational tasks such as configuration generation, troubleshooting, and quantitative reasoning.
Yet they remain mostly language-level evaluations, while benchmarks for wireless world models are still missing.
Even in this limited setting, strong frontier models struggle with schema-compliant intent-to-configuration, supporting GSMA's conclusion that agentic architectures are needed.
This aligns with our position: deployable telecom AI requires explicit interfaces between foundation models and specialized solvers, as well as meaningful benchmarks.
We provide a broader discussion of related work in \refapp{app:related_work}.
\section{Structural Bottlenecks for Wireless AI Data}
\label{sec:bottlenecks}

Current approaches to wireless LLMs and world models are difficult to reconcile with established training methods because they face structural data limitations.
Wireless data is often underspecified, highly context-dependent, and difficult to universally tokenize, making it unclear whether standard scaling laws~\citep{kaplan2020scaling,hoffmann2022training} transfer to this domain.
As a result, both LLMs and world models risk offering limited deployable value unless data interfaces, operational scope, and feedback mechanisms are explicitly defined.
We identify four main bottlenecks:

\paragraph{Bottleneck 1: Configuration Dependence.}
In wireless settings, a data sample is inseparable from the system configuration that produced it.
The same learning problem changes entirely under different carrier frequencies, antenna geometries, beam codebooks, pilot designs, numerologies, and schedulers.
Thus, a CSI tensor labeled \emph{``good channel state''} under one configuration is meaningless under another, and training across all plausible configurations is combinatorially infeasible.
This leaves wireless AI models with three unpalatable options:
(i)~\emph{infer the configuration from the signal}, which works in-distribution but fails silently outside it,
(ii)~\emph{condition explicitly on configuration metadata}, which is rarely done systematically or even available, or
(iii)~\emph{average over a mix of configurations}, yielding models that are mediocre everywhere and reliable nowhere.
3GPP's focus on concrete air-interface use cases reflects precisely this constraint and aims to isolate tasks to make their data regime tractable.
Thus, a wireless dataset without configuration metadata is a dataset for interpolation, not intelligence.

\paragraph{Bottleneck 2: Simulation and Setup Dependence.}
Wireless AI relies heavily on simulation because real deployment data is scarce, proprietary, and difficult to label.
Platforms such as DeepMIMO~\citep{alkhateeb2019deepmimo}, Sionna~\citep{hoydis2022sionna}, OpenRAN Gym~\citep{bonati2023openran}, and Colosseum~\citep{polese2024colosseum} provide essential infrastructure for data generation, but every simulator also encodes its own assumptions about signal propagation, interference, and protocol abstractions.
Thus, models trained on one simulator may learn its artifacts rather than the underlying wireless physics, echoing the sim-to-real problem in robotics.
Thus, unlike LLMs, they cannot be easily downloaded and reused across arbitrary configurations, simulators, and deployments.
Wireless AI therefore requires explicit setup provenance, including the simulator name and version, scenario, channel model, configuration, impairments, and random seeds.

\paragraph{Bottleneck 3: No Universal Wireless Token.}
LLMs and world models are successful in encoding vast knowledge in part because text tokens are broadly reusable abstractions.
However, IQ samples, CSI tensors, beam indices, and scheduler traces differ not only in format, but also in abstraction level, control timescale, and causal regime. 
\emph{A CSI tensor is not a sentence. An IQ stream is not a paragraph. A scheduler trace is not source code.}
Moreover, the task context often lives outside the sample, e.g., a CSI tensor alone says little about antenna geometry, pilot structure, feedback protocol, or mobility model.
Thus, before asking how to train a wireless foundation model, one must answer: \emph{What is the wireless equivalent of a token, instruction, and tool call?}
Although recent works attempt to tokenize or unify wireless streams~\citep{liu2024llm4cpadaptinglargelanguage,djuhera2026mambacsphybridattentionstatespace}, these representations remain task- and modality-specific.
This heterogeneity is significant: PHY receivers, MAC schedulers, and RAN optimizers operate at different timescales, with different losses, action spaces, latency constraints, and safety requirements.
The wireless stack is modular for a reason and AI should respect those boundaries unless deployment evidence justifies breaking them.

\paragraph{Bottleneck 4: Missing Operational Feedback.}
Modern LLMs improve through instruction tuning, retrieval feedback, and preference optimization~\citep{ouyang2022training,lewis2020retrieval}.
However, post-training alignment as understood for LLMs has no wireless counterpart, because there is no scalable notion of operator or user preference over PHY/MAC/RAN actions, and no safe way to learn one from live trial and error. 
Operators cannot freely A/B test schedulers on live users, standards and vendor constraints limit admissible actions, and the most informative failures are precisely those too costly to induce.
World models do not resolve this either and only relocate the problem, since the model itself must be calibrated against operational ground truth data that the same missing feedback channel is supposed to provide.
Thus, our position remains that the only architecture compatible with the operational data output of wireless networks is an agentic one, in which reasoning models consume structured data (telemetry, KPIs, configuration logs) through explicit interfaces rather than learning from a preference signal that does not exist.

Given these bottlenecks, it is evident that wireless datasets should be accompanied by data cards that document configuration, provenance, and other assumptions (see \refapp{app:wireless_data_problem}).

\begin{figure*}[t]
    \centering
    \includegraphics[width=0.9\linewidth]{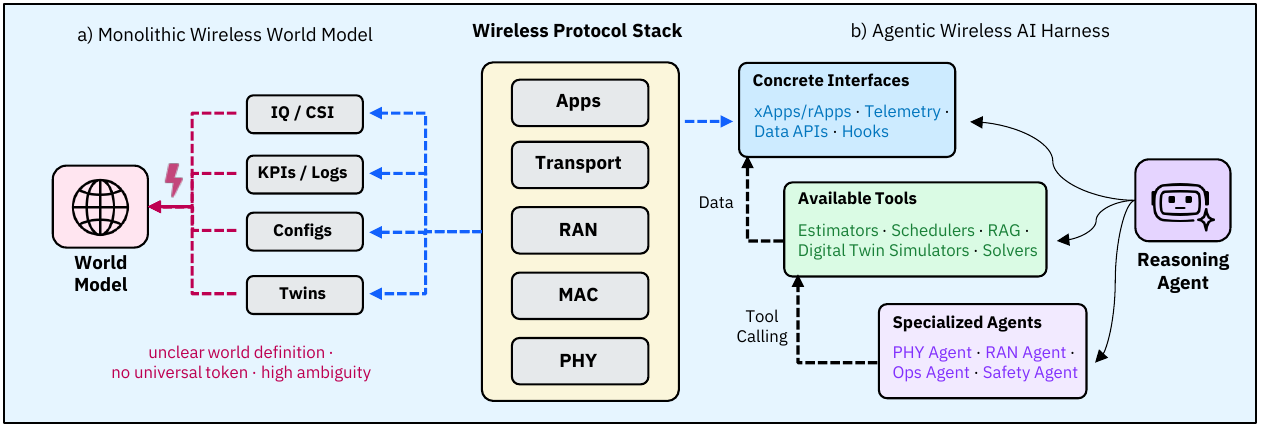}
    \caption{Monolithic wireless world models collapse heterogeneous data into an ambiguous interface, whereas agent harnesses orchestrate tools, interfaces, and specialized agents through explicit and auditable tool calls. See \refapp{app:agentic_wireless_network_architectures} for a detailed architecture description.}
    \label{fig:agentic_vs_world_model}
\end{figure*}
\section{The Case Against Wireless World Models}
\label{sec:world_models}

A wireless world model is a learned simulator of radio network dynamics (channels, queues, KPIs), conditioned on operator actions such as beam updates, handover policies, or resource allocations (see \refapp{app:world_models}).
It is intended to support counterfactual rollouts (e.g., \emph{``What if we changed the scheduler?"}) without interacting with the live network.
This makes world models attractive for sample-efficient and proactive planning that classical reinforcement learning (RL) cannot offer.
However, we argue that the structural data bottlenecks are inherited in world models and that they are compelling only when the ``world'' is explicitly scoped.
Indeed, recent proposals already point toward an agentic composition rather than monolithic pretraining~\citep{zou2026telecom}.
We make the following claims:

\paragraph{The Wireless World is Not One World.}
In the wireless domain, the relevant ``world'' changes with the control problem.
For beam tracking, it may consist of channel states, mobility, blockage, and beam codebooks over millisecond horizons.
For network slicing, it may consist of queues, PRB allocations, transport congestion, and SLA margins over seconds or minutes.
These are not merely different features of one homogeneous state space, but correspond to different abstraction levels, timescales, action spaces, and objective functions.
A single world model that tries to absorb all of them must either flatten these distinctions into one opaque latent state or hand-design interfaces between them.
The former makes the model difficult to interpret and validate, while the latter already turns the system into a modular architecture.
Thus, the problem is not that wireless worlds cannot be defined, but that useful definitions are tied to local predictive models for particular control loops, defying the notion of a single wireless world model.
Moreover, such models inherit rather than resolve the data bottlenecks: action-conditioned rollouts require logged trajectories, spatial prediction requires configuration-aware RF data, synthetic pretraining requires simulator provenance, and cross-layer modeling still lacks a shared tokenization.
World models therefore make all four bottlenecks load-bearing at once.

\paragraph{World Models Become Agentic When Made Practical.}
World models may be valuable for traffic generation, digital-twin acceleration, and counterfactual planning~\citep{zheng2026digitaltwinsworldmodelsopportunities}, but they should not be treated as default replacements for robust channel estimators, schedulers, or operator workflows.
Recent work~\citep{zou2026telecom} acknowledges this and proposes to decompose wireless world models into three layers: 
(i)~\emph{a Field World Model layer} for spatial prediction, 
(ii)~\emph{a Control World Model layer} for action-conditioned KPI rollouts, and 
(iii)~\emph{a Foundation Model layer} for intent translation and orchestration. 
The latter is explicitly agentic: it coordinates calls to world-model components, simulators, digital twins, O-RAN xApps/rApps, policy engines, and constraint solvers, while validating candidate actions against SLA and safety constraints before execution.
Thus, the most developed case for wireless world models is not one for monolithic intelligence, but for compositional, tool-using architectures compatible with existing telecom infrastructure, thereby undermining initial definitions of a single, unitary wireless world model.

The strongest argument \emph{for} wireless world models therefore becomes an argument against treating them as the endpoint: they motivate an agentic architecture in which world-model components are tools, not the system itself.
\section{A Composable and Agentic Alternative}
\label{sec:agentic}

If monolithic models are the wrong abstraction, the alternative is not to abandon them, but to place them inside an \emph{agent harness}, a runtime layer in which reasoning agents orchestrate specialized wireless models, classical algorithms, simulators, telemetry databases, retrieval systems, and safety monitors through explicit interfaces, rather than replacing the protocol stack (see \refig{fig:agentic_vs_world_model}).
This architecture is more aligned with how wireless systems are already built.

Specifically, PHY, MAC, RAN, transport, core, and management layers are highly modular and O-RAN further exposes this modularity through near-real-time and non-real-time control loops, xApps/rApps, telemetry streams, and policy interfaces~\citep{bonati2023openran}, providing the necessary connectors to agentic workflows~\citep{dev2025advanced}.
Thus, PHY tasks can remain with specialized signal-processing models, RAN optimization can be handled by scoped xApps/rApps, digital twins can provide counterfactual validation, and standards retrieval can constrain configuration logic.
Within the harness, LLM-based reasoning agents are equipped with tool-use capabilities~\citep{masterman2024landscape} and translate intents, plan multi-step workflows, select tools, and escalate uncertain cases to human operators.
The agentic layer thus enables guardrailed orchestration that monolithic world models cannot without becoming modular themselves.

Furthermore, the agentic approach also directly addresses the wireless data bottlenecks in \refsec{sec:bottlenecks}.
First, configuration dependence is handled by making configuration an explicit input to each tool call rather than an implicit variable hidden inside a latent model.
Second, simulation dependence is made auditable because simulator identity, scenario, channel model, and random seed can be attached to every generated rollout.
Third, the lack of a universal wireless token is avoided: CSI tensors, IQ streams, KPI logs, topology graphs, and standards documents need not be forced into one representation, but can remain in their native formats behind specialized interfaces.
Fourth, missing operational feedback is mitigated by separating recommendation from execution. 
The corresponding agent harness can run shadow-mode analyses, compare actions in digital twins, enforce guardrails, and request operator approval before affecting live users.
In addition, new agent skills can be added modularly as use cases evolve and as future 3GPP releases introduce new interfaces, procedures, and requirements.

For example, consider a cell experiencing degraded beam-management performance under high mobility.
An agentic system first retrieves the cell configuration, beam codebook, UE mobility traces, and recent KPI degradation.
It then calls a classical beam-selection baseline and a learned beam predictor, compares their confidence and latency, and tests both under nearby mobility scenarios in a digital twin.
If the learned predictor is confident and passes the simulator and safety checks, the agent can recommend switching the corresponding xApp to the learned model.
Thus, wireless modules are not replaced globally or permanently, but selected dynamically according to the environment.
Further examples and detailed architectures are provided in \refapp{app:agentic_wireless_network_architectures}.

Consequently, intelligence arises from composition rather than from forcing one model to internalize the entire wireless stack.
This architecture also remains adaptable because individual models, tools, and workflows can be replaced, reconfigured, or validated independently as network conditions, standards, and deployment requirements evolve.
\section{Conclusion}
\label{sec:conclusion}

In this paper, we argued that monolithic wireless world models face structural data bottlenecks that scale alone cannot resolve, thereby limiting their broad usability.
As an alternative, we advocate composable AI-native wireless systems, in which agent harnesses coordinate specialized wireless models and classical solvers through explicit interfaces.
Our position is thus clear: AI-native wireless should not be measured by whether the community can train ever larger models, but whether wireless intelligence can be made deployable, configuration-aware, and, most importantly, compatible with the modular protocol stack.


\section*{Acknowledgments}

This work was supported in part by the German Federal Ministry of Research, Technology and Space (BMFTR) within the research hubs 6G-life (Grant 16KISK002) and 6GEM+ (Grant 16KIS2409K) through the projects GEM-X (Grant 16KISS004K), VICTOR6G (Grant 16KIS2547), QD-CamNetz (Grant 16KISQ077), and AISAC (Grant 16KIS2462). This work was also supported by the Bavarian Ministry of Science and the Arts through the project Next Generation AI Computing (gAIn), by the Bavarian Ministry of Economic Affairs, Regional Development and Energy through the project 6G Future Lab Bavaria, by the German Research Foundation (DFG) within the Centre for Tactile Internet with Human-in-the-Loop (CeTI) as part of Germany’s Cluster of Excellence (EXC 2050/2, ID 390696704), and in part by IBM Research.
\section*{Impact Statement}

This paper presents work whose goal is to advance the field of machine learning and wireless communications. 
There are many potential societal consequences of our work, none of which we feel must be specifically highlighted here.


\bibliography{bibliography}
\bibliographystyle{icml2026}


\appendix
\section{Related Work}
\label{app:related_work}

Recent progress in AI has accelerated research across the wireless protocol stack.
This includes both \emph{AI for wireless}, where learning methods improve communication, sensing, and network-control tasks, and \emph{wireless for AI}, where communication systems support distributed training, inference, and edge deployment of AI models.
We briefly summarize the most relevant directions, with emphasis on foundation models, LLMs, agents, and world models.

\subsection{AI for Wireless and Wireless for AI}
Early AI-for-wireless research focused on well-scoped optimization and inference problems. 
Neural and reinforcement-learning methods have been applied to channel estimation~\citep{jin2019channel}, beam prediction~\citep{li2023machine}, resource allocation~\citep{djigal2022machine}, mesh networks~\citep{karunaratne2019overview}, trajectory optimization~\citep{cheng2021machine}, and sensing~\citep{he2023robust}.
These works demonstrated early that learning can be valuable when the task interface, input representation, and evaluation metric are clearly specified.

More recently, the rise of LLMs and attention-based architectures has motivated their adaptation to wireless sequence modeling.
For example, \citet{liu2024llm4cpadaptinglargelanguage} and \citet{djuhera2026mambacsphybridattentionstatespace} adapt sequence models to channel-state prediction by designing CSI tokenization pipelines over time, frequency, and delay-domain representations.
Beyond channel prediction, \citet{zhang2025wi} propose an LLM-powered Wi-Fi sensing system that maps raw CSI into textual or visual representations. 
Similarly, \citet{wang2026beamagent} propose an LLM-aided MIMO beamforming framework in which the LLM parses natural-language intent into structured spatial constraints, while a dedicated numerical optimizer performs joint site selection and precoding.
Furthermore, \citet{zou2026rfgptteachingaiwireless} introduce RF-GPT, an RF perception interface which converts IQ signals into spectrograms and grounds them through synthetic metadata and instruction tuning.
These works show the potential of LLM-style interfaces, but they also reinforce a recurring pattern: successful systems either require task-specific tokenization or decouple the LLM from the underlying numerical wireless computation.

A complementary line of work studies \emph{wireless for AI}, where communication systems are designed to support distributed learning and inference.
This includes resilient federated and split-learning mechanisms under jamming, noise, or unreliable wireless links~\citep{R_SFLLM,R_MTLLMF}, as well as joint model partitioning, scheduling, and inference placement across edge servers, MEC nodes, and on-device LLM services~\citep{djuhera2026joint,zhang2025communication,liu2025willm}.
These efforts highlight that AI-native wireless is not only about applying AI to the network, but also about adapting the network to the computational and communication demands of modern AI workloads.

More recently, related trends seek to reproduce the cross-task capabilities of LLMs through telecom-specific post-training.
For example, \citet{TSpecLLM,TeleLLMs,TeleQnA} develop telecom-oriented datasets and models for standards understanding, Q\&A, retrieval, and domain-specific reasoning.
These efforts are important first steps, but they mostly train on textual artifacts of telecom knowledge rather than on the operational substrate of wireless systems.
They therefore improve language-level expertise, but do not by themselves ground models in RF measurements, network dynamics, control actions, or deployment feedback.
To bridge this gap, recent work has proposed large wireless foundation models, multimodal wireless models, and wireless world models~\citep{alikhani2024lwm,LMMs,zou2026telecom,saad2025agi}.
However, many of these proposals remain at the level of architectural vision, with limited discussion of the data interfaces, provenance, operational feedback, and evaluation protocols required for deployment.
This raises the question of whether current foundation and world model visions adequately address the structural data bottlenecks discussed in \refsec{sec:bottlenecks}.

Overall, existing work validates the usefulness of both classical neural methods and attention-based models for wireless applications.
However, most successes remain tied to scoped tasks, carefully designed representations, and explicit optimization or system interfaces.
This motivates our central position: deployable AI-native wireless should compose specialized models, solvers, simulators, and reasoning agents, rather than assuming that a single monolithic model can absorb the entire protocol stack.

\subsection{Simulators, Datasets, and Benchmarks}
As AI-based methods become increasingly common in wireless research, high-quality simulators and experimental platforms are essential for data generation, training, testing, and reproducible evaluation.
Platforms such as DeepMIMO~\citep{alkhateeb2019deepmimo}, Sionna~\citep{hoydis2022sionna}, OpenRAN Gym~\citep{bonati2023openran}, and Colosseum~\citep{polese2024colosseum} provide important infrastructure for simulating wireless environments, signal propagation, protocol behavior, and network-control loops, often with interfaces to modern AI frameworks such as PyTorch~\citep{paszke2019pytorch}.
These platforms have been instrumental for developing and validating AI-based wireless models.
At the same time, they also encode assumptions about channel models, propagation, traffic, mobility, hardware impairments, and protocol abstractions.
As a result, models trained in one simulation environment may learn simulator-specific artifacts rather than transferable wireless structure.
This motivates systematic cross-simulator and sim-to-real studies, as well as stronger documentation of simulator provenance, calibration, and standards compliance.

Beyond simulators, recent works have introduced telecom-oriented datasets for LLM post-training and evaluation~\citep{TSpecLLM,TeleLLMs,TeleQnA}.
As discussed above, these datasets are useful for standards understanding, Q\&A, and retrieval, but remain predominantly text-based and therefore provide limited grounding for wireless use cases.
They also inherit common risks of post-training, including safety degradation and domain overfitting.
For instance, \citet{djuhera2025safecomm} show that telecom post-training can degrade safety behavior and introduce TeleHarm as a telecom-specific safety realignment dataset.
These findings suggest that telecom datasets must move beyond textual knowledge artifacts toward multi-modal, configuration-aware, and operationally grounded data.

Benchmarking remains similarly underdeveloped.
Current telecom LLM benchmarks, including the benchmarks proposed by~\citet{gsma_open_telco_2025}, mostly measure standards knowledge, Q\&A, configuration generation, or troubleshooting, but do not yet capture closed-loop control quality, causal intervention accuracy, robustness under distribution shift, or deployment safety.
This raises a central question: \emph{How should actionable wireless intelligence be measured?}
A credible benchmark suite would need to evaluate not only language-level correctness, but also whether a system can retrieve the right state, choose the right tool, respect protocol and safety constraints, and improve network KPIs under realistic operational conditions.
Without such benchmarks, wireless world models may remain compelling architectural visions, but their practical value for deployable network intelligence cannot be meaningfully assessed.

While a complete survey is beyond the scope of this paper, existing trends show that both classical neural methods and attention-based models are increasingly used to improve wireless applications.
However, as argued in the main body, most works remain tied to scoped tasks, custom representations, and isolated benchmarks, leaving open the question of how these components should be composed into deployable, auditable, and configuration-aware AI-native wireless systems that are compatible with existing telecom infrastructure, protocol stacks, and standards.
\section{The Wireless Data Problem and Data Cards}
\label{app:wireless_data_problem}

Wireless systems generate enormous amounts of data, but much of it is difficult to reuse for general-purpose AI.
Compared with text, code, or images, an IQ stream, CSI tensor, KPI trace, or scheduler log is meaningful only relative to its configuration, protocol state, measurement setup, simulator assumptions, and deployment context.
In \refsec{sec:bottlenecks}, we identified four structural bottlenecks that make monolithic wireless models, including wireless world models, difficult to develop and validate.
In particular, the lack of a universal documentation scheme prevents wireless data from becoming a reusable foundation-model substrate.

This motivates our composable and agentic view of AI-native NextG: rather than forcing all wireless data into a single model, specialized components should consume well-documented data through explicit interfaces compatible with existing protocol and standardization boundaries.
However, agentic wireless AI still requires a substantial redesign of how wireless datasets are categorized, documented, and shared.
We therefore propose \emph{wireless data cards}: structured documentation artifacts that record the configuration, provenance, task, and deployment assumptions behind each dataset.
This is inspired by dataset and model documentation practices in responsible ML, but adapted to the particular needs of wireless systems.

A minimal wireless data card should record:

\begin{itemize}[leftmargin=1em,itemsep=0pt,topsep=2pt]
\item \textbf{Task:} prediction target, action space, loss function, latency budget, and safety constraints.
\item \textbf{Configuration:} carrier frequency, bandwidth, numerology, antenna geometry, beam codebook, pilot design, and scheduler settings.
\item \textbf{Provenance:} measurement source or simulator, simulator version, channel model, geometry, mobility and traffic models, impairment assumptions, and random seeds.
\item \textbf{Environment:} indoor/outdoor setting, topology, blockage, deployment density, UE distribution, and mobility.
\item \textbf{Protocol Layer and Timescale:} PHY, MAC, RAN, core, edge, application layer and interaction frequency.
\item \textbf{Policy:} operator objectives such as fairness, energy, QoS/QoE, as well as regulatory and safety constraints.
\item \textbf{Evaluation:} baselines, distribution shifts, failure criteria, uncertainty calibration, and robustness tests.
\end{itemize}

Without such documentation, wireless datasets remain configuration-opaque and difficult to compose across tasks, simulators, and deployments.
For agentic systems, data cards are even more important: they allow an agent to decide which dataset, simulator, model, or tool is valid for a given configuration and control objective.
Future extensions may also include reasoning traces, tool-call histories, simulator rollouts, and operator feedback, enabling agentic workflows to be audited and improved over time.
The call for wireless data cards is therefore a call for wireless AI to adopt the same rigor in data documentation that modern AI already demands, while accounting for the configuration- and deployment-dependence unique to wireless systems.
\section{Wireless World Models}
\label{app:world_models}

In this section, we formalize the notion of a wireless world model and review related work.

\subsection{Definition and Taxonomy}
World models are appealing because they promise prediction before action.
In their classical form, a world model learns an action-conditioned transition model,
\begin{equation}
    p_\theta(s_{t+1:t+H}\mid s_t,a_{t:t+H}),
\end{equation}
where \(s_t\) denotes the current latent or observed system state, \(a_{t:t+H}\) denotes a candidate action sequence over a planning horizon \(H\), and \(s_{t+1:t+H}\) denotes the predicted future trajectory under that action sequence.
Such a model allows an agent to imagine possible futures, evaluate candidate actions, and plan under uncertainty.
However, in the LeCun sense~\citep{assran2023self}, a world model is useful only when the ``world'', the state representation, the action space, and the prediction targets are well defined.
For wireless systems, this immediately raises a specification problem.
Is the state an electromagnetic field, a CSI tensor, a queue vector, a traffic map, a slice state, a protocol trace, an operator policy, or all of these jointly?
Are the actions beam updates, scheduling decisions, handover parameters, routing rules, slice budgets, energy-saving policies, or application-level intents?
What is the prediction horizon: microseconds, slots, seconds, minutes, or hours?
What objective should be optimized: throughput, latency, reliability, energy, fairness, SLA compliance, or some operator-specific trade-off?
Without specifying these elements, a wireless world model is not yet a model, but only a metaphor.

\subsection{Recent Work on Wireless World Models}
Recent wireless and telecom world model proposals are most convincing precisely when they avoid ambiguity.

Initial work on world model-based learning for long-term age-of-information (AoI) minimization defines a concrete mmWave V2X link-scheduling problem, a packet-completeness-aware AoI objective, a recurrent state-space model (RSSM) for latent dynamics prediction, and a Sionna-based simulator with ray tracing and vehicle mobility~\citep{wang2025worldmodelbasedlearninglongterm}.
The follow-up dual-mind wireless world model~\citep{wang2025dmwmdualmindworldmodel} further specializes this setting by adding a pattern-driven System~1 component and a logic-driven System~2 component, aiming to improve long-horizon imagination and scheduling consistency under mobility, blockage, and link-availability constraints.
This line of work is valuable because the ``world'' is carefully scoped: the state contains packet-completeness-aware AoI (CAoI), vehicle locations, and channel/ray-tracing features, the action is link scheduling, and the objective is long-horizon CAoI minimization.
It is therefore a useful local world model, but not evidence that a single model can absorb the entire wireless stack.

At the network level, \citet{zou2026telecom} provide one of the most explicit attempts to define a telecom world model (TWM).
Rather than treating the world as an undifferentiated wireless latent space, TWM decomposes the telecom system into a controllable system world and an exogenous world.
It then introduces three layers: a \emph{Field World Model} for spatial and electromagnetic prediction, a \emph{Control/Dynamics World Model} for action-conditioned KPI trajectories, and a \emph{TelecomGPT layer} for intent translation, tool orchestration, and guardrail enforcement.
This formulation explicitly clarifies what a telecom world model would need to provide: state grounding, action-conditioned rollouts, uncertainty estimates, multi-timescale dynamics, model-based planning, and safety-aware interaction with operators.
However, it also reinforces our main argument.
The strongest TWM formulation is not monolithic, but decomposed, tool-integrated, simulator-dependent, and explicitly agentic at the foundation model layer.

In addition, recent work connecting digital twins and world models~\citep{zheng2026digitaltwinsworldmodelsopportunities} argues that learned predictive models can accelerate simulation, support counterfactual planning, and enable mobile edge or network intelligence beyond reactive control.
We agree with this direction when world models are treated as scoped predictive components.
Digital twins provide high-fidelity mirroring, validation, and data generation, while world models may provide faster learned rollouts for specific control loops.
The problem arises when this complementarity is replaced by the stronger claim that a single wireless world model should become the central substrate for AI-native networking.

\subsection{Structural Data Bottlenecks Remain}
As argued in the main body, wireless world models inherit the same data bottlenecks as wireless LLMs, but in a stricter form.
A world model must predict transitions under actions.
Missing context therefore does not merely reduce accuracy, but invalidates counterfactual rollouts.
Configuration dependence becomes a transition-model problem: the same CSI tensor, beam update, or scheduler action can imply different future dynamics under different numerologies, antenna layouts, traffic models, or beam codebooks.
Simulation dependence becomes a calibration problem: if rollouts are generated by a simulator with biased propagation, mobility, or protocol assumptions, the world model may learn simulator artifacts as if they were wireless physics.
The absence of a universal wireless token becomes an interface problem: IQ samples, CSI tensors, topology graphs, KPI logs, and operator intents cannot be fused by simply scaling a single sequence model without destroying useful abstraction boundaries.
Finally, missing operational feedback becomes a causal-identifiability problem: historical telemetry reflects actions chosen by existing operator policies, not the counterfactual outcomes of unseen interventions.
Thus, world models do not bypass the wireless data problem and make configuration metadata, simulator provenance, cross-layer interfaces, and operational calibration even more load-bearing.

Our argument is therefore not that wireless world models are useless.
Local world models can be useful for scoped prediction problems: radio-map forecasting, traffic evolution, slicing rollouts, failure propagation, link scheduling, or counterfactual policy evaluation.
The problem is the stronger claim that monolithic or single wireless world models should become the primary substrate of AI-native networking.
In the wireless domain, the relevant world is fragmented across PHY, MAC, RAN, transport, core, edge, and application layers, each with different observations, actions, timescales, constraints, and safety requirements.
The more realistic path is to treat world models as specialized predictive tools inside a broader agentic architecture, alongside classical algorithms, RF perception models, digital twins, telemetry databases, standards-aware retrieval, and safety monitors.
World models are thus useful only when the world is sufficiently scoped, especially in wireless which is a hierarchy of coupled worlds operating at incompatible timescales.
\section{Agentic Wireless Network Architectures}
\label{app:agentic_wireless_network_architectures}

In this section, we expand on the agentic wireless network architecture proposed in \refsec{sec:agentic} and summarize why the limitations of current wireless LLMs and world models point toward such composable architectures.
Furthermore, we outline how agentic reasoning can be integrated with existing telecom infrastructure, protocol stacks, and workflows.

\subsection{Why the Evidence Points to Agentic Architectures}
The case for agentic wireless AI follows directly from the structural bottlenecks discussed in \refsec{sec:bottlenecks}.
Most importantly, the lack of a universal wireless token actively prohibits training truly multimodal wireless foundation models. 
Accordingly, monolithic models cannot resolve these bottlenecks, and we argue that only agentic systems can overcome them.

For example, a reasoning agent can query configuration databases, retrieve standards constraints, call specialized PHY/MAC/RAN models, invoke classical solvers, run digital twin validations, and check safety policies before recommending an action.
However, this does not imply that AI-based methods should replace existing wireless algorithms.
Many signal processing, estimation, optimization, and scheduling methods remain preferable in regimes where they are reliable, interpretable, and computationally efficient.
The role of the agent is therefore not to replace the wireless stack, but to select, compose, and validate the right component for the current task, configuration, and risk level.

This view is also consistent with how telecom systems are standardized and deployed.
3GPP and O-RAN do not expose the network as one end-to-end differentiable system, but instead define functions, interfaces, measurements, control loops, and constraints.
Agentic wireless AI treats this modularity as a design advantage, whereas monolithic models tend to obscure it.
Rather than asking one model to internalize the entire protocol stack, agentic introduces intelligence at the interface level, where reasoning, retrieval, tool use, validation, and human oversight can be made explicit.

\subsection{Architecture: Agents Around the Protocol Stack}
\begin{figure}[t]
    \centering
    \includegraphics[width=0.95\linewidth]{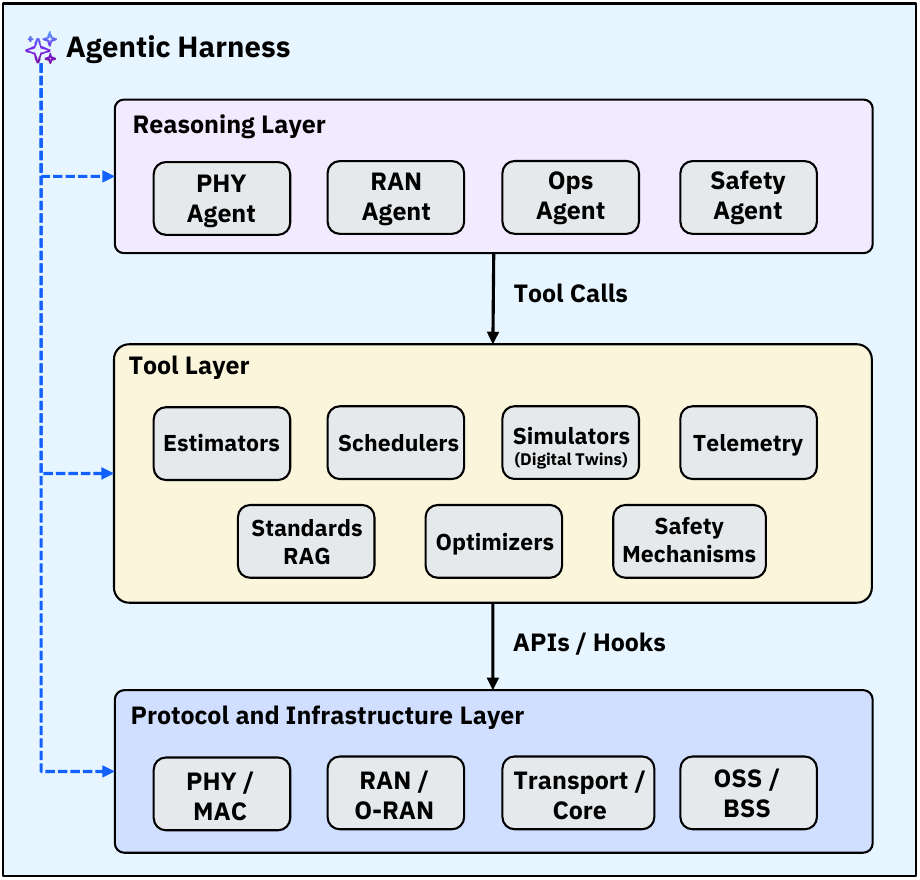}
    \caption{Three-layer agent harness for wireless networks.}
    \label{fig:agentic_three_layers}
    \vspace{-1em}
\end{figure}

An agentic wireless network architecture consists of three layers connected by an \emph{agent harness} (see \refig{fig:agentic_three_layers}).
The first layer is the \emph{protocol and infrastructure layer}, including PHY/MAC procedures, RAN control, transport/core functions, telemetry pipelines, configuration databases, and existing management systems.
The second layer is the \emph{tool layer}, exposing specialized components through callable interfaces: channel estimators, beam predictors, interference classifiers, schedulers, optimizers, simulators, digital twins, standards retrieval, and safety monitors.
The third layer is the \emph{reasoning layer}, where one or more agents plan workflows, select tools, compare outputs, validate constraints, and communicate with operators.
The harness connects these layers by managing context, permissions, tool schemas, logging, guardrails, rollback, and human approval.

In a practical deployment, multiple agents can operate at different scopes and timescales.
A PHY-facing agent may assist with model selection, channel estimator validation, or beam management diagnosis.
A RAN-control agent may coordinate xApps/rApps for load balancing, energy saving, slicing, or mobility optimization.
A safety agent can check proposed actions against SLA, regulatory, vendor, and rollback constraints.
The agents coordinate through structured tool calls and shared state summaries, while latency-critical decisions remain inside specialized control loops.

This architecture is compatible with current infrastructure.
O-RAN already separates control across near-real-time and non-real-time timescales and exposes modular xApps/rApps, telemetry streams, and policy interfaces.
Containerized deployment further allows operators to add, update, or roll back individual models and tools without retraining a full-stack model.
Agent skills can be defined as bounded capabilities, such as \texttt{query\_kpi}, \texttt{retrieve\_3gpp\_constraint}, or \texttt{compare\_estimators}.
This makes the system highly maintainable, as operators can audit which tools were called, which assumptions were used, which constraints were checked, and why a recommendation was made.

\subsection{Example Workflows}
We sketch three representative workflows that illustrate how agentic wireless AI composes existing components.

\paragraph{Model Selection for Channel Estimation.}
A gNB observes degraded CSI quality for a subset of UEs under changing mobility and SNR conditions.
The agent retrieves the carrier configuration, pilot pattern, antenna geometry, UE mobility profile, recent CSI error statistics, and historical channel estimation performance.
It then compares a classical LS/MMSE estimator, a lightweight neural estimator, and a larger learned CSI predictor under the current configuration.
The candidates are evaluated on recent telemetry and, when needed, on simulated channel realizations with matched numerology and mobility.
If the neural estimator improves NMSE while satisfying latency and uncertainty constraints, the agent recommends enabling it.
If the scenario is outside the learned model's documented validity range, the agent falls back to the classical estimator and flags the need for additional data.
Thus, the estimator is selected per configuration and risk level, not replaced globally.

\paragraph{Scheduler Adaptation Under Slice Pressure.}
A cell experiences rising URLLC delay while eMBB throughput remains high.
The agent retrieves per-slice queues, PRB utilization, HARQ statistics, traffic forecasts, current scheduler weights, and SLA constraints.
It then calls a queueing-based baseline, a learned scheduler surrogate, and a digital twin rollout to compare candidate scheduling policies.
Unsafe policies are rejected if they violate URLLC latency, starve eMBB traffic, or exceed predefined fairness limits.
The agent may recommend a bounded update to scheduler weights, a temporary slice-priority adjustment, or no change if the evidence indicates transient congestion.
The key point is that the agent does not invent a scheduler from scratch.
It selects and validates among existing scheduling mechanisms using telemetry, prediction, and policy constraints.

\paragraph{Intent-to-Configuration for Network Operations.}
An operator issues a high-level intent such as \emph{``Prepare this region for a stadium event while preserving URLLC reliability and limiting energy cost.''}
The agent retrieves the relevant cells, neighbor relations, traffic history, mobility forecasts, slice policies, and applicable 3GPP/O-RAN constraints.
It decomposes the intent into candidate actions, such as temporary slice budget changes, mobility parameter updates, admission control thresholds, or activation of additional carriers.
Each candidate is checked against standards constraints, evaluated through a digital twin or learned surrogate, and passed through a safety monitor before being surfaced to the operator.
The final output is not a raw configuration dump, but a ranked plan with assumptions, expected KPI impact, uncertainty, and rollback conditions.
This illustrates the appropriate role of foundation models for translating intent and orchestrating tools, while validated components perform the wireless computation.

These examples illustrate the key architectural principle: intelligence arises from composition.
The agent does not hallucinate network states when telemetry can be queried, does not reinvent protocol behavior when standards can be retrieved, and does not replace a validated solver when a solver is the right tool.
Such workflows can be learned from wireless reasoning traces, which are already implicit in current network operations through telemetry queries, configuration changes, troubleshooting tickets, simulator runs, and operator decisions.
Agentic wireless AI is therefore not a rejection of foundation models, world models, or classical algorithms.
It is a deployment-oriented framework for deciding when each of them should be used.


\end{document}